\documentstyle[prl,aps,epsf,floats]{revtex}
\begin{document}
\draft
\twocolumn[\hsize\textwidth\columnwidth\hsize\csname @twocolumnfalse\endcsname
\title {Density-Induced Breaking of Pairs in the Attractive
Hubbard Model}
\author{Bumsoo Kyung,$^1$ E.\,G.\,Klepfish,$^2$ 
and P.\,E.\,Kornilovitch$^1$}
\address{$^1$Max-Planck-Institut f\"ur Physik komplexer Systeme, 
N\"othnitzer str. 38, D-01187, Dresden, Germany\\
$^2$Department of Physics, King's College London, Strand,
London, WC2R 2LS, UK}
\date{5 September 1997}
\maketitle
\begin{abstract}

   A conserving $T$-matrix approximation is applied to the two-dimensional
attractive Hubbard model in the low-density regime. 
A set of self-consistent equations
is solved in the real-frequency domain to avoid the analytic 
continuation procedure. By tuning the chemical potential
the particle density was varied in the limits $0.01 < n < 0.18$.
For the value of the attractive potential $U=8 \, t$
the binding energy of pairs monotonically decreases
with increasing $n$, 
from its zero-density limit $2.3 \, t$ and vanishes at a
critical density $n_{cr}=0.19$. A pairing-induced pseudogap in the
single-particle density of states is found at low densities
and temperatures.
\end{abstract}
\pacs{PACS numbers: 71.10.Fd, 74.20.Mn}
\vskip2pc]
\narrowtext

   The pseudogap state of underdoped cuprates is currently
one of the central topics in the studies of high-temperature
superconductivity (HTSC). The pseudogap (PG) is believed to
be responsible for marked deviations from Fermi-liquid
behavior observed in a variety of probes. Indications
from NMR \cite{Takigawa,Williams_one}, specific heat
measurements \cite{Loram}, transport \cite{Ito,Batlogg},
optics \cite{Rotter,Homes}  
were followed by more recent angle-resolved photoemission 
\cite{Loeser}, Raman \cite{Nemetschek} 
and core-electron photoemission \cite{Tjernberg} studies.
A consensus seems to be emerging that the PG is of
crucial importance for understanding HTSC because it is
intimately related to superconductivity. The PG has the same
symmetry as the superconducting gap 
\cite{Loeser,Nemetschek,Williams}, it decreases
with hole doping \cite{Loram,Batlogg,Puchkov} 
and has a universal doping dependence when measured in units 
of the maximum critical temperature \cite{Williams}.

   The PG state gets a simple explanation in phenomenological
models which assume an effective short-range attractive
interaction between holes doped into a Mott-Hubbard
insulator \cite{Alexandrov,Randeria}. 
PG is then associated with the binding
energy $\Delta$ of hole pairs. Superconductivity sets
in at lower temperatures 
$T_c \sim n_{pairs}/m_{pairs}$ \cite{Uemura}.
Thus, at small hole densities $\Delta$ and $T_c$ are
well separated. Microscopic origin of such an attraction
may be different varying from pure electronic \cite{Dagotto}
to pure phononic \cite{Alexandrov} mechanisms, see also
\cite{Ranninger}.

   In this Letter we address the doping dependence of
the PG. It has already been discussed by Alexandrov, Kabanov,
and Mott in \cite{Alexandrov_two}. They found that the
binding energy of pairs decreases with doping as a
result of the screening of the long-range electron-phonon
interaction and renormalization of the inter-hole coupling.
Here we will show that PG decreases with doping even
in the case of short-range inter-hole interaction and 
without coupling renormalization. This is a pure many-body
effect which is expected on the following grounds. 
If the coupling exceeds the binding threshold then, 
at low density, the carriers form
non-overlapping pairs. As the density
increases pairs begin to overlap and fermions, which 
constitute pairs, begin to form a Fermi sea. 
This process increases the energy of the system per particle
hence raising the energy of the two-particle
level. Under certain conditions this level could be 
pushed up to the single-particle continuum. Near the
critical density the binding energy and the life-time
of pairs become small and pairs break up. 

   In order to verify this idea we study the two-dimensional
attractive Hubbard model which is the simplest fermion
lattice model with attraction. The model is defined by
Hamiltonian
\begin{equation}
H = \sum_{{\bf k}\sigma} (\varepsilon_{\bf k} -\mu)
c^{\dagger}_{{\bf k}\sigma} c_{{\bf k}\sigma} -
\frac{|U|}{N} \sum_{\bf k p q}
c^{\dagger}_{{\bf k} \uparrow} c_{{\bf k+q} \uparrow}
c^{\dagger}_{{\bf p} \downarrow} c_{{\bf p-q} \downarrow}
\label{one}
\end{equation}
written in the standard notation. 
$\varepsilon_{\bf k}=-2t \, (\cos{k_x}+\cos{k_y})$ is the
single-particle bare spectrum, $|U|$ is the coupling strength,
$N$ is the total number of sites in the system and chemical potential
$\mu$ determines the average particle density. We will be interested 
in the low-density regime of the model. In this 
regime one can make use of the small gas parameter and
select only ladder diagrams in the diagrammatic representation
of the Bethe-Salpeter equation \cite{Galitskii,Kadanoff}. 
This procedure leads to the $T$-matrix approximation for
the Hubbard
model \cite{Kanamori} which we formulate here in the real-frequency
domain. The single-particle spectral function $A({\bf k}, \omega)$
is determined by the real and imaginary parts of the self-energy
$\Sigma'$ and $\Sigma''$
\begin{equation}
A({\bf k}, \omega) = \frac{-\, 2\, \Sigma''({\bf k}, \omega)}
{[\omega - (\varepsilon_{\bf k}-\mu)-\Sigma'({\bf k}, \omega)]^2
+ [\Sigma''({\bf k}, \omega)]^2} .
\label{two}
\end{equation}
The self-energy is expressed, in turn, via the real and imaginary
parts $T'$ and $T''$ of the $T$-matrix
\begin{displaymath}
\Sigma'({\bf k}, \omega) = \frac{1}{N} \sum_{\bf q}
\int \frac{d \omega_1}{2\pi} A({\bf q-k}, \omega_1) \times
\end{displaymath}
\begin{equation}
\times \left[ f_F(\omega_1) T'({\bf q}, \omega+\omega_1) +
\int \frac{d \omega_2}{\pi}
\frac{f_B(\omega_2) T''({\bf q}, \omega_2)}{\omega_2-\omega_1-\omega}
\right]
\label{three}
\end{equation}
\begin{displaymath}
\Sigma''({\bf k}, \omega) = \frac{1}{N} \sum_{\bf q}
\int \frac{d \omega_1}{2\pi} A({\bf q-k}, \omega_1) \times
\makebox[2.cm]{}
\end{displaymath}
\vspace{-0.3cm}
\begin{equation}
\times T''({\bf q}, \omega+\omega_1)
\left[ f_F(\omega_1) + f_B(\omega+\omega_1) \right]
\label{four}
\end{equation}
where $f_{F,B}(\omega)=[\exp(\beta \omega) \pm 1]^{-1}$ 
are Fermi- and Bose- functions respectively, and $\beta=1/T$ is the
inverse absolute temperature. Finally, the $T$-matrix is expressed via
the spectral function as follows
\begin{equation}
T({\bf q}, \omega) = \frac{-|U|} {1-|U| \displaystyle{
\int \frac{d\omega_1}{2\pi} 
\frac{B({\bf q}, \omega_1)}{\omega-\omega_1} + 
i \frac{|U|}{2} B({\bf q}, \omega) }}        
\label{five}
\end{equation}
\begin{equation}
B({\bf q}, \omega) \!=\! \frac{-1}{N} \! \sum_{\bf k'}
\!\! \int \!\! \frac{d\omega_1}{2\pi} A({\bf k'}\!, \omega_1)
A({\bf q\!-\!k'}\!, \omega\!-\!\omega_1)  \tanh{\!\frac{\beta
\omega_1}{2}}.
\label{six}
\end{equation}
The integrals with singular kernels in Eqs.(\ref{three}) and
(\ref{five}) are understood in the principal value sense.
The set of equations (\ref{two})-(\ref{six}) is to be solved
self-consistently for given $|U|$, $\mu$ and temperature
$T$, the particle density given by:
\begin{equation}
\frac{n}{2} = \frac{1}{N} \sum_{\bf k} 
\int^{\infty}_{-\infty} \frac{d\omega}{2\pi} 
A({\bf k},\omega) f_F(\omega) . 
\label{seven}
\end{equation}
Single-particle energies are associated with sharp peaks
in the spectral function $A({\bf k},\omega)$. Analogously, the energies
of two-particle bound states can be inferred from peaks of
the imaginary part of the $T$-matrix $T''({\bf q}, \omega)$
which is proportional to the spectral function of the two-particle
propagator with total quasi-momentum ${\bf q}$, 
$A_2 ({\bf q}, \omega) = - T''({\bf q}, \omega)/|U|$. 
The difference between the pair energy and the doubled
single-particle energies, particularly the dependence
of this difference on the density of particles, 
is the main object of this work.

   The method of numerical solution of the set of equations
is well-known: one has to iterate Eqs. (\ref{two})-(\ref{six})
starting with some initial guess for the spectral function
and to use the Fast Fourier Transform (FFT) algorithm to
speed up the calculation of the integrals. We note
that in previous analogous studies \cite{Wolfle,Micnas,Engelbrecht} 
the equations were formulated in the imaginary-time domain which
required an analytic continuation of the iteration results
to real frequencies. This ill-posed problem of numerical
analytic continuation leads to ambiguities in the 
details of reconstructed functions. Therefore we prefer 
to deal directly with the real-frequency equations. 
We have observed that the singular
structure of the kernels in Eqs. (\ref{two})-(\ref{six})
is not an obstacle for the FFT-integration and does not slow
down the convergence of the iterative process.     

   Our calculations were done on a $64 \times 64$-site
lattice and on a uniform mesh of 512 points in the frequency
interval $-20\,t < \omega < 30\,t$.
Eqs. (\ref{two})-(\ref{six}) were iterated until 
the relative difference between two successive iterations
fell below 0.01 (usually 20-25 iterations). Note, that {\em no}
artificial damping factor was used in the iteration process.
The resulting spectral functions were checked against 
the normalization property 
$\int \frac{d\omega}{2\pi} A({\bf k},\omega) = 1$ (for other
sum rules for the Hubbard model see e.g. 
\cite{Nolting,Micnas}) which were found
to be satisfied within 5\% accuracy for each ${\bf k}$
(normalization was accurate within 1\%). The 
temperature was taken $T=1.0 \, t$.
\begin{figure}[t]
\begin{center}
\leavevmode
\hbox{
\epsfxsize=8.6cm
\epsffile{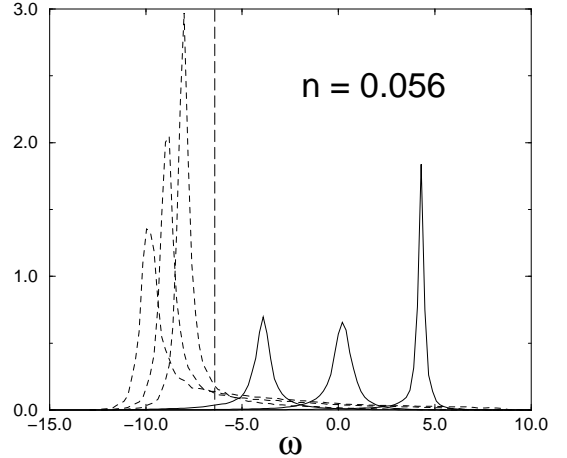}
}
\end{center}
\vspace{-0.5cm}
\caption{
Spectral function $A({\bf k},\omega)/(2\pi)$ (solid lines)
and $A_2({\bf k},\omega)=-T''({\bf k},\omega)/|U|$
(dashed lines) for three values of
lattice momentum, ${\bf k}=(0,0)$, $(\pi/2,\pi/2)$ and
$(\pi,\pi)$ (from left to right). Energy is measured with respect
to the on-site single-particle level i.e. $A({\bf k},\omega)$ are
shifted by $\mu$ and $A_2({\bf k},\omega)$ by $2\,\mu$ from
results of self-consistent solution. The position of the
chemical potential $\mu=-6.4 \,t$ is shown by the vertical
long-dashed line. $|U| = 8\,t$, temperature
$T=1.0 \,t$ and the corresponding particle density is $n=0.056$.
}
\label{fig1}
\end{figure}

\begin{figure}[t]
\begin{center}
\leavevmode
\hbox{
\epsfxsize=8.6cm
\epsffile{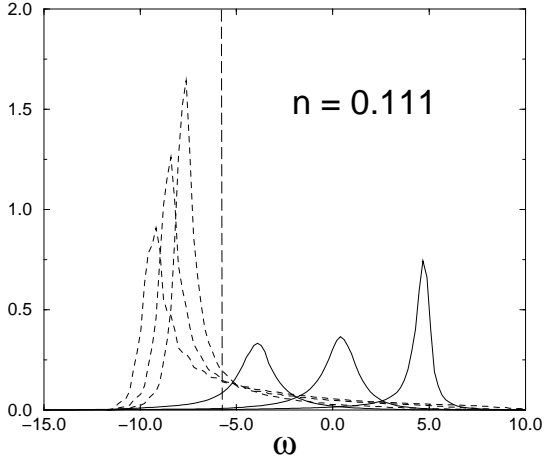}
}
\end{center}
\vspace{-0.5cm}
\caption{
The same as Fig. 1 but for chemical potential $\mu=-5.8 \,t$
corresponding to a larger density $n=0.111$.
}
\label{fig2}
\end{figure}

   Figs. \ref{fig1} and \ref{fig2} show results of iterations
for two different values of the chemical potential
$\mu=-6.4 \,t$ and $\mu=-5.8 \,t$ corresponding to particle densities
$n=0.056$ and $n=0.111$ respectively. To obtain the binding
energy of a bound state with total quasi-momentum ${\bf q}$
one has to compare the position of the corresponding peak in
$A_2({\bf q},\omega)$ with the doubled position of a peak
in spectral function $A({\bf k},\omega)$ taken at half
momentum $({\bf q}/2)$. Thus, in Fig. \ref{fig1} the binding
energy of the pair with ${\bf q}=(0,0)$ is 
$2 \, E_1(0,0) - E_2(0,0) \approx 2.3 \, t$, the one of
the pair with ${\bf q}=(\pi,\pi)$ is 
$2 \, E_1(\pi/2,\pi/2) - E_2(\pi,\pi) \approx 8.0 \, t$.  
An analysis of Figs. \ref{fig1} and \ref{fig2} and analogous
data for seven other densities in the region $0.01 \leq n \leq 0.18$
(not shown) reveals the following common trends. 
i) The single-particle peak as a function of momentum
${\bf k}$ spans an energy interval of width $\approx 8 \, t$
for all the densities studied. This implies small effect
of the interaction on the shape and width of the single-particle
energy spectrum. The same conclusion has been reached in
Ref.\cite{Micnas}. ii) The lower boundary of the single-particle
band $E_1(0,0)$ shifts towards negative   
values as density increases. However, the shift observed is smaller 
than the mean-field value $-|U|n/2$. Apparently, this is due to
the correlation effects. At $|U|=8\,t$ the effective radius of
a pair is $\approx 1 - 2$ lattice spacings, see e.g. 
\cite{Kornilovitch}. Therefore the interaction of a 
particle with spin $\sigma$ excited to the continuum spectrum, 
with a bound particle of spin $-\sigma$ is effectively screened
by the latter's spin $\sigma$ partner.  
iii) The energy of the two-particle state {\em increases}
with the density, as indeed was expected on the grounds of
the qualitative considerations. iv) The single particle and the
pair spectral functions become narrower at large lattice 
momenta. The life-time of the bound states with large momenta  
increases which results from reduced phase space of the 
scattering process into two single-particle states.  
v) The binding energy of the pair with the highest momentum
${\bf q}=(\pi,\pi)$ is always close to $|U|$ (which is equal
to $8 \, t$ in our case). This result of a fully self-consistent
treatment is in perfect agreement with the exact solution
of the two-body problem (the zero-density limit) \cite{Note_one}
and the general theory of the $\eta$-resonance \cite{Demler}
according to which a collective mode at 
${\bf q}=(\pi,\pi)$ with binding
energy $|U|$ always exists in the attractive Hubbard model.

\begin{figure}[t]
\begin{center}
\leavevmode
\hbox{
\epsfxsize=8.6cm
\epsffile{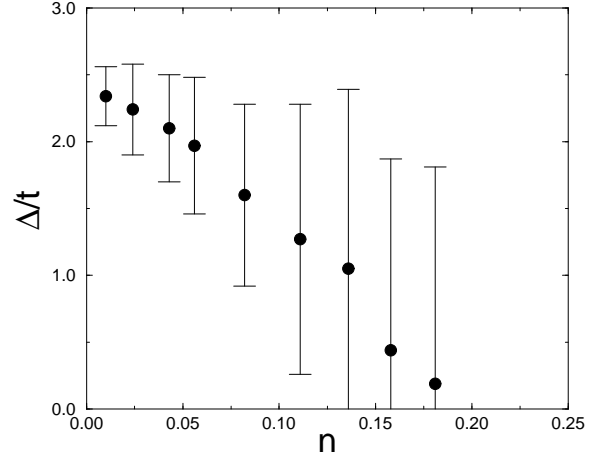}
}
\end{center}
\vspace{-0.5cm}
\caption{
Pair binding energy vs. particle density. The points
are determined from the position of the maxima of peaks
of spectral functions. Bars indicate the width of peaks.
}
\label{fig3}
\end{figure}

   At low density only states with small momenta are occupied.
Therefore we now concentrate on the properties of
pairs with ${\bf q}=(0,0)$. We have analysed the 
doping-dependence of the binding
energy using solutions of the self-consistent equations
for nine different densities in the interval  
$0.01 \leq n \leq 0.18$. For the smallest density
studied, $n=0.01$, the binding energy was found to be 
$\approx 2.3 \, t$. For the two-particle problem
the binding energy in the ground state is $\Delta (n=0)=2.1\,t$
\cite{Note_one}. The proximity
of these two values confirms the correctness of our numerical
treatment of the self-consistent equations. The small difference
is apparently due to the finite-size effects.  
The density dependence of the binding energy is shown
in Fig.\ref{fig3}. 
It is a monotonically 
decreasing function which vanishes at $n_{cr} \approx 0.19$.
The points have been calculated from the positions
of the maxima of one- and two-particle spectral functions.
The peaks' widths are associated with
finite life-time of the excitations. The bars in Fig.\ref{fig3} 
indicate the width (the inverse life-time) of the 
two-particle spectral weight. Obviously, if the energy
gap between the bound states and the single-particle
continuum diminishes the scattering process of the pair
into two single-particle states intensifies. This should
lead to decrease of the life-time and widening of 
the peaks.
This is exactly what is seen in the self-consistent results.
We have checked that the decrease of the binding energy 
is to be attributed mainly to the increase of 
the energy of two-particle states.
There is also a contribution from the negative shift
of single-particle continuum, but we have found this
process to account for $\approx 10 \%$ of the whole
effect. It is also clear that the region of 
stability of the bound states depends on the coupling
constant: it expands for larger $|U|$'s and shrinks for
smaller ones. We have repeated all the calculations
for $|U|=4.0\,t$ and $T=0.2\,t$ and found that for density
$n=0.10$ the pair with ${\bf q}=(0,0)$ is not a well-defined 
excitation, its binding energy and life-time being too small. 
The pair excitation separates from the single-particle
continuum at ${\bf q} \approx (\pi/2,\pi/2)$. At the
corner of the Brillouin zone the pair's binding
energy is $\approx |U|=4.0\,t$. 
Therefore, the coupling is to be increased if one
requires well-defined pair excitations at ${\bf q}=(0,0)$.
At large couplings the convergence of the iterative process
worsens but restores with increasing temperature.
Thus, we were able to trace the destruction of pairs
in full at $|U|=8.0\,t$ and $T=1.0\,t$. 

\begin{figure}[t]
\begin{center}
\leavevmode
\hbox{
\epsfxsize=8.6cm
\epsffile{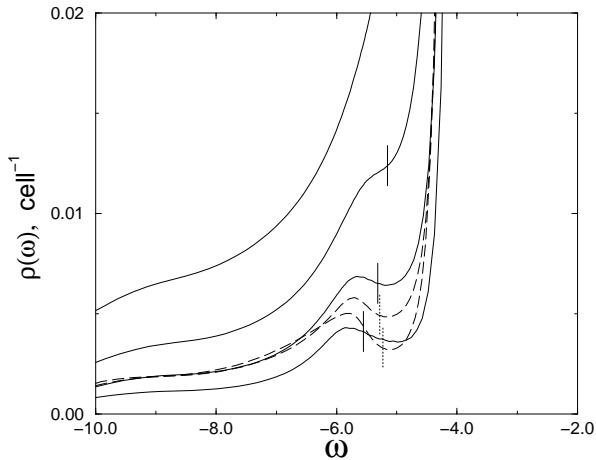}
}
\end{center}
\vspace{-0.5cm}
\caption{
The low-energy part of the density of states for
$|U|=8\,t$. Solid lines are for $T=0.5\,t$ and densities $n=
1.0, 1.7, 3.3$ and $6.3\%$ (from the lowest curve upwards).
Dashed lines are $n=1.7\%$ and $T=0.4\,t$ (upper curve) and
$T=0.3\,t$. Vertical lines mark
the energies $\omega_0$ such that $\int^{\omega_0}_{-\infty}
d\omega \rho(\omega) = n$.
}
\label{fig4}
\end{figure}

   The density dependence of the pair binding energy
found in our calculations is similar to that 
of the pseudogap observed in the underdoped regime of high-$T_c$
superconductors. This supports the real-space pairing
as the cause for the pseudogap phenomenon. In order to
test and clarify this relation further we calculated
the single-particle density of states $\rho(\omega)$ from 
the solution of the $T$-matrix equations. The low-energy part 
of $\rho(\omega)$ is shown in Fig. \ref{fig4}. 
At small 
$n$ we observe an extra peak in $\rho(\omega)$
which is apparently due to the presence of bound states,
since its separation from the main part of $\rho(\omega)$
is approximately the binding energy. (Compare with the
atomic limit of the Hubbard model, see e.g. \cite{Micnas}.)
At larger
$n$ the feature shifts to higher energies and finally
disappears thereby qualitatively repeating the $n$-dependence
of the binding energy discussed above. Two comments are
to be added. First, the density of states is an integral 
of the single-particle spectral function over momenta 
${\bf k}$. Since at different ${\bf k}$ the additional maxima  
appear at different energies the feature is less pronounced
in $\rho(\omega)$ than in individual $A({\bf k},\omega)$. 
Second, upon lowering the temperature the local minimum in
$\rho(\omega)$ deepens, see the dashed curves in Fig. \ref{fig4}.
This implies the opening of a pseudogap at low temperatures,
as experimentally observed in the underdoped cuprates.
Note that the position of the peak is not affected by
temperature which confirms that it is the pair binding 
energy. 

   In conclusion, we have studied the low-density regime
of the two-dimensional attractive Hubbard model by means
of the self-consistent $T$-matrix approximation.
The set of self-consistent equations was solved
in the real-frequency domain to avoid uncertainties
related to the analytic continuation procedure.
The binding energy of pairs was found to be a monotonically
decreasing function of the total density. We interprete 
this as a pure many-body effect resulting 
from the rise of two-particle levels to the single-particle
continuum due to the packing effect. We believe 
this effect is generic to low-density regimes of fermionic
models with attraction and it does not depend on the
dimensionality or particular details of the interaction.
In particular it should remain in more realistic models
of low-density hole systems such as the
nearest-neighbors attraction model with $d$-symmetrical
bound states. We observe also a pairing-induced
pseudogap in the density of states with the characteristic
size equal to the binding energy of pairs. The pseudogap
develops at low densities upon lowering temperature.

   We acknowledge valuable discussions on the subject
with A.\,S.\,Alexandrov, A.\,M.\,Dyugaev, S.\,Flach, P.\,Fulde, 
A.\,O.\,Gogolin, R.\,B.\,Laughlin, A.\,A.\,Nersesyan, 
Yu.\,N.\,Ovchinnikov, and E.\,R.\,Pike. EGK has been supported by EPSRC
grant GR/J18675.
\end{document}